\documentclass[twocolumn,pre,numerical,superscriptaddress,longbibliography]{revtex4-1}
\usepackage{graphicx,color,epsfig,subfig,dcolumn,bm,mathrsfs,amsmath}
\usepackage[colorlinks]{hyperref}

\newcommand{\red}[1]{\textcolor{black}{#1}}
\newcommand{\fs}[1]{\textcolor{black}{#1}}

\begin{document}

\title{Complex Formation between Polyelectrolytes and 
Oppositely Charged Oligoelectrolytes}

\author{Jiajia Zhou}
\email[]{zhou@uni-mainz.de}
\affiliation{School of Chemistry \& Environment, Center of Soft Matter Physics and Its Applications \\
Beihang University, Xueyuan Road 37, Beijing 100191, China}
\affiliation{Komet 331, Institute of Physics, Johannes Gutenberg-University Mainz \\
Staudingerweg 9, D55099 Mainz, Germany} 
\author{Matthias Barz}
\affiliation{Institute of Organic Chemistry, Johannes Gutenberg-University Mainz \\
Duesbergweg 10-14, D55099 Mainz, Germany} 
\author{Friederike Schmid}
\email[]{friederike.schmid@uni-mainz.de}
\affiliation{Komet 331, Institute of Physics, Johannes Gutenberg-University Mainz \\
Staudingerweg 9, D55099 Mainz, Germany} 

\date{\today}

\begin{abstract}
We study the complex formation between one long polyanion chain and many short
oligocation chains by computer simulations.  We employ a coarse-grained
bead-spring model for the polyelectrolyte chains, and model explicitly the
small salt ions. We systematically vary the concentration and the length of the
oligocation, and examine how the oligocations affects the chain conformation,
the static structure factor, the radial and axial distribution of various
charged species, and the number of bound ions in the complex.
At low oligocation concentration, the polyanion has an extended structure.  Upon
increasing the oligocation concentration, the polyanion chain collapses and
forms a compact globule, but the complex still carries a net negative
charge. Once the total charge of the oligocations is equal to that of the
polyanion, the collapse stops and is replaced by a slow expansion. In this
regime, the net charge on the complexes is positive or neutral, depending on
the microion concentration in solution. The expansion can be explained by the
reduction of the oligocation bridging. We find that the behavior and the
structure of the complex are largely independent of the length of oligocations,
and very similar to that observed when replacing the oligocations by
multivalent salt cations, and conclude that the main driving force keeping the
complex together is the release of monovalent counterions and coions.  We speculate
on the implications of this finding for the problem of controlled oligolyte
release and oligolyte substitution.
\end{abstract}

\maketitle

\section{Introduction}
\label{sec:introduction}

Polyelectrolytes are linear macromolecules composed of ionizable groups
\cite{Oosawa}.  Many important macromolecules in biology are polyelectrolytes,
such as DNA and proteins.  Polyelectrolytes are usually water soluble
and are therefore widely used in water-based organic formulations,
which are of great technological interest due to their economical and
environmental benefits.  
When dissolved in water, the ionizable groups dissociate into small mobile ions
and leave the chain backbone with the opposite charge. The electrostatic
interaction strongly influences the chain conformations.  Due to the
interplay of short-range excluded volume and long-range electrostatic
interactions, and the presence of the small ions, polyelectrolyte solutions
display a rich variety of intriguing phenomena, and their behavior is much less
understood than that of neutral polymer solutions.  Therefore, they are also a
rewarding subject for theoretical and simulation studies
\cite{Barrat1996, Dobrynin2005}.

Polyelectrolytes can be categorized into weak polyelectrolytes or strong
polyelectrolytes, depending on the fraction of charged monomers. In strong
polyelectrolytes, the Coulomb repulsion between charged monomers forces the
chains to expand, and their resulting radius of gyration is much larger than
that of a neutral Gaussian chain of the same length.  Many applications of
polyelectrolyte solutions depend on the possible control of the chain
conformations.  For example, the rheology of a polyelectrolyte solution changes
dramatically when the polyelectrolyte conformations change from extended
structures to compact globules \cite{Fuoss1951, Dobrynin1995}.  The resulting
rapid variation in the viscosity of the solution has found many industrial
applications, such as the automobile brake system.  Another notable example is
non-viral gene delivery, where a therapeutic DNA or RNA is transferred into
specific cells of patients \cite{Pack2005, Glover2005}.  The genetic materials
are required to pass many obstacles, to penetrate the cell and nuclear
membranes.  These tasks can only be performed efficiently when the size of the
delivery vector is a few tens to several hundreds nanometers.  In addition,
\fs{the morphology of the complex} is expected to influence its biological functions
\cite{Hayashi2016}.  Therefore, the conditions of complex formation will affect
\fs{the} complex properties and finally the transfection efficiency.

Polyelectrolyte complexes are aggregates of polyelectrolyte chains and
oppositely charged species.  Taking polyanions as an example, one can
choose many positively charged complexing agents.  The most common choice
are multivalent salts. The effect of adding
multivalent salt cations to polyanion solutions
is well documented in the literature \cite{Bloomfield1997}.  
Upon increasing the multivalent salt
concentration, the initially homogeneous polyelectrolyte solution becomes phase
separated if the salt concentration is higher than a critical value.  This
critical value is proportional to the polyelectrolyte concentration, and
essentially corresponds to the value at which the total charge of added
multivalent cations neutralizes the total charge of the polyanions.  Above this
critical concentration, the solution demixes into a dense and viscous
precipitate phase and a dilute solution phase.  This phase separation is
associated with complex formation between the polyanion and multivalent
cations, and the complexes aggregate to form the precipitate phase.  Upon
further increase of the salt concentration, the precipitate dissolves, and
the solution transforms back into a homogeneous phase.  The reentrance
behavior to the disordered phase can also be associated \fs{with} the reexpansion of
the polyanion chain.

Another popular choice of complexing agent is a polyelectrolyte
with opposite charge, especially in the field of non-viral gene delivery
system \cite{Wagner1990, Miyata2012}. Polymeric transfection systems have been
under rapid development for the last two decades, and \emph{in vivo}
applications also start to emerge \cite{Laechelt2015}.  Because of the flexibility of the polymer
chemistry, polyelectrolytes can be synthesized in linear, branched, and
dendritic structures.  It is also possible to design them such that they
contain not only charged monomers, but also include neutral or hydrophobic
blocks that may provide multiple functions. One can tailor the
complexing polyelectrolytes to fulfill the specific requirements for
efficient delivery, which makes polyelectrolytes a popular choice for
gene-delivery vectors.

Despite the advance of experimental techniques, the resolution of single
polyelectrolyte complexes remains difficult.  Simulation studies can help
to understand the organization of the complex in molecular detail,
if appropriated models are used.  They also provide insight into the
underlying physics. For example, it is now well understood that complex
formation in strong polyelectrolytes is driven by the gain of
translational entropy associated with the release of counterions, rather
than by the gain of enthalpy due to the electrostatic interactions
\cite{OuZhaoyang2006}.  Simulations can also help to explore the parameter
space, to save the time and money for trial-and-error experiments.  So far,
most simulation studies have focused on multivalent cations \cite{Liu2002,
Liu2003, OuZhaoyang2005, HsiaoPai-Yi2006, HsiaoPai-Yi2006a, HsiaoPai-Yi2006b},
or on the complexation of oppositely charged polyelectrolytes with similar
length \cite{Srivastava1994, Winkler2002, Hayashi2002, Hayashi2003,
Hayashi2004, Dias2003, OuZhaoyang2006}. 
%
%

In this work, we study the complex formation of one long
polyelectrolyte chain with many short oppositely charged
oligoelectrolytes. We systematically vary the concentration of the short
chains and analyze the structure and composition of the complex.  We use
Langevin simulations to study the complex formation.  The remainder of this
article is organized as follows: In section \ref{sec:model}, we briefly
introduce the simulation model and describe important parameters of the
system.  We present the simulation results on the complex structure in section
\ref{sec:results}. Finally, we conclude in section \ref{sec:summary} with a
brief summary.

\section{Simulation model}
\label{sec:model}

We employ a coarse-grained model \cite{Stevens1995} to simulate the flexible
polyelectrolyte chains. Our systems contain two types of polyelectrolyte
chains which we will label as A-chain and B-chain for better reference:
An A-chain is a negatively charged polyelectrolyte (a polyanion)
composed of $N_{A}=100$ beads, and a B-chain is a positively charged
oligomer (an oligocation) composed of $N_{B}$ beads ($N_B$ in the range of 2--5).  We keep the number of
A-chain fixed at $n_A=1$ in the simulation box, and we vary the number of
B-chain $n_{B}$.  Throughout this work, we will characterize the
systems in terms of the monomer concentrations, which are given by
\begin{equation} 
  \rho_A = N_A / V, \quad \quad \rho_B = n_B N_B / V,
\end{equation} 
where $V$ is the volume of the simulation box.  Both A- and
B-chains are uniformly charged; each bead carries one charge unit.
For each polyelectrolyte, a corresponding number of oppositely charged
monovalent ions (i.e., $N_A$ cations and $n_B N_B$ anions) is added to the
system to make the whole system charge neutral. We shall refer to the
cations as counterions (with respect to the A-chain), and to the anion as
coions.  In Table \ref{tab:beads}, we list all variables to describe different
species in the system.

\begin{table}[htbp]
\centering
\begin{tabular}{| p{2.5cm} || c | c | c | c |}
\hline
type             & length  & charge & molecule & monomer \\ 
                 &         &        & number     & number  \\ \hline \hline
A-chain          & $N_{A}$ & $-1$ & 1             & $N_A$     \\ \hline
B-chain          & $N_{B}$ & $+1$ & $n_{B}$       & $n_B N_B$ \\ \hline
Counterion ($+$) & 1       & $+1$ & $N_{A}$       & $N_A$     \\ \hline
Coion ($-$)      & 1       & $-1$ & $n_{B} N_{B}$ & $n_B N_B$ \\ \hline
\end{tabular}
\caption{Table of simulation variables.}
\label{tab:beads}
\end{table}

Short-range excluded volume interactions between each pair of beads 
are modeled by the repulsive part of the Lennard-Jones interaction
(a Weeks-Chandler-Anderson potential \cite{WCA}).
\begin{equation}
\label{eq:LJ}
  U_{\rm LJ} (r) = \begin{cases} 
    4 \varepsilon \left[ \left(\frac{\sigma}{r}\right)^{12} 
    - \left(\frac{\sigma}{r}\right)^6 + \frac{1}{4} \right] 
    & r< \sqrt[6]{2} \sigma \\
    0 & {\rm otherwise} 
  \end{cases}
\end{equation}
where $r$ is the distance between two beads, and $\varepsilon$ and $\sigma$
characterize the energy and length scales. 
Here the shift of the potential level
is chosen such that the potential is zero at the cutoff. 

The beads in the polyelectrolyte chains are connected by harmonic springs.
This bonded interaction has the form
\begin{equation}
\label{eq:harm}
  U_{\rm harm} (r) = \frac{1}{2} K ( r-r_b )^2 ,
\end{equation}
where $r$ is the distance between two connected beads, $K$ is the spring
constant, and $r_b$ is the equilibrium bond length.  The parameters are chosen as
$K = 5000\, \varepsilon/\sigma^{-2}$ and $r_b=\sigma$.  For these choices, the bond length
fluctuates within $10\%$ of the equilibrium bond length \cite{Liu2002}, and
chain crossing is prohibited.

Charged beads also interact with each other {\em via} electrostatic interaction. 
For monovalent charged beads, the Coulomb interaction can be written as
\begin{equation}
\label{eq:coul}
  U_{\rm Coul}=\frac{ e^2 }{ 4 \pi \epsilon \epsilon_0 r} = k_BT \frac{ \ell_B }{r}.
\end{equation}
Here $\epsilon$ is the dielectric constant of the medium and $\epsilon_0$ is
the vacuum permittivity, $k_BT$ is the Boltzmann constant times the
temperature.  The Bjerrum length characterizes the distance at which the
electrostatic potential of a pair of unit charges becomes comparable to the thermal
energy, $\ell_B = e^2/4\pi\epsilon\epsilon_0 k_BT$.  In this study, we focus
on strongly charged polyelectrolytes and choose $\ell_B = 3.0\, \sigma$.
Electrostatic interactions are calculated using the standard P3M method
\cite{HockneyEastwood, Deserno1998}.

We perform Langevin simulations on the complex formation, where water is
treated implicitly.  The effect of the water is incorporated
{\em via} a viscous environment that provides a coupling to a thermal bath in
the equations of motions of the beads, and a dielectric background for the
Coulomb interaction.  The equation of motion for $i$-th bead is taken as
\begin{equation}
  \label{eq:Langevin}
  m \frac{ \mathrm{d}^2 \mathbf{r}_i } { \mathrm{d} t^2 } 
    = - \xi \mathbf{v}_i - \mathbf{\nabla}_{\mathbf{r}_i} U + \mathbf{f}_i(t),
\end{equation}
where $m$ and $\xi$ are the monomer mass and friction coefficient,
respectively. Equation (\ref{eq:Langevin}) implies that we neglect hydrodynamic
interactions, which is acceptable because we are mostly interested in static
equilibrium properties of the system.  $U$ is the total potential energy
consisting of the Lennard-Jones interaction (\ref{eq:LJ}), the
harmonic spring interaction (\ref{eq:harm}), and the electrostatic interaction
(\ref{eq:coul}).  The term $\mathbf{f}_i(t)$ refers to a random force
that mimicks the effect of thermal motion due to the surrounding solvents.
This noise term satisfies the fluctuation-dissipation theorem
\red{
\begin{equation}
  \langle {f}_{\alpha i}(t) \cdot {f}_{\beta j} (t') \rangle 
    =  2k_BT \xi \, \delta_{\alpha \beta} \delta_{ij} \delta(t-t'),
\end{equation}
where $\alpha,\beta=x,y,z$ indicating the components of the random force.}
We have taken the mass of all beads as unit mass.  The temperature of the
system is set at $k_BT=1.0\, \varepsilon$.  The friction coefficient is
chosen $\xi=1.0\, \tau^{-1}$, where $\tau=\sigma \sqrt{m/\varepsilon}$ is
the time unit of the simulation.  We have used the velocity-Verlet scheme
\cite{Swope1982, FrenkelSmit} to integrate Eq.~(\ref{eq:Langevin}), with a time
step $0.01\, \tau$.  All simulations are carried out using the open-source
package ESPResSo \cite{ESPResSo}.

We consider cubic simulation boxes with two box sizes: one is $100\,
\sigma$ and the other one is $25\, \sigma$.  The corresponding A-monomer
concentrations are $10^{-4}\, \sigma^{-3}$ and $6.4\times 10^{-3}\,
\sigma^{-3}$, respectively.  Compared to the size of a freely-jointed
chain of $100$ monomers, $R_e \sim 10\, \sigma$, the corresponding
systems are still in the dilute regime for the A-chain. The main
effect of the system size is to regulate the density of counterions and
coions, as will be discussed below.

We start the simulation with the A-chain in the center of the box, and
randomly distributed B-chains.  We then perform the Langevin dynamics
simulation until the system reaches equilibrium.  This equilibration process
normally takes $2\times 10^6$ time steps.  After equilibration, we take a
snapshot and record the position of each bead every $2000$ steps in the
following $2\times 10^6$ time steps. 
\red{Three runs with independent starting configurations are performed for each parameter setting.} 
These results are used to perform
a statistical analysis and compute the physical quantities discussed in the 
next section.

In some cases, we also perform reference simulations with multivalent
salt ions. Here we replace the B-chains composed of $N_B$ monovalent
charged beads by single beads carrying $N_B$ unit charges.  The
comparison between the oligocation case and the multivalent salt case can
provide insight into the influence of the chain character of oligocations
on the structure of the polyelectrolyte complexes.

\section{Results and Discussions}
\label{sec:results}

\subsection{Single chain conformation}
\label{sec:Rg}

The size and the shape of a polymer chain in solution can be characterized by
several quantities.  In this section, we calculate these quantities for the
polyanion chain and study their dependence on the B-monomer concentration.

Specifically, we compute the time-average of the following quantities:
\begin{enumerate}
\item {\em The radius of gyration.}
We monitor the gyration tensor of the A-chain, defined by
\begin{equation}
  \mathbf{S} = \frac{1}{N} \sum_{i=1}^N (\mathbf{r}_i - \mathbf{r}_{\rm com}) 
                \otimes (\mathbf{r}_i - \mathbf{r}_{\rm com}),
\end{equation}
where $\mathbf{r}_{\rm com}$ is the center-of-mass position.  The gyration
tensor can be written as a symmetric $3\times3$ matrix, which can be
diagonalized by a proper rotation. We denote the diagonal elements of the
$\mathbf{S}$-matrix by $\lambda_1^2$, $\lambda_2^2$, and $\lambda_3^2$, and
without loss of generality, we assume $\lambda_1^2 \ge \lambda_2^2 \ge
\lambda_3^2$.  The radius of gyration is given by 
\begin{equation}
  R_g = \sqrt{ \lambda_1^2 + \lambda_2^2 + \lambda_3^2 }.
\end{equation}

\item {\em The end-to-end distance.}
The end-to-end distance is defined as 
\begin{equation}
  R_e = | \mathbf{r}_1 - \mathbf{r}_{N} | ,
\end{equation}
where $\mathbf{r}_1$ and $\mathbf{r}_{N}$ is the position of the first and the last beads of A-chain. 
\red{The ratio $R_e^2/R_g^2$ gives some information about the shape conformation. 
This ratio increases from a value of 6 (for a Gaussian chain) to 12 (for a rodlike structure).}

\item {\em The hydrodynamic radius.}
The hydrodynamic radius, sometimes also called Stokes radius,
characterizes the dynamic properties of the whole chain moving in the
solvent.  It can be computed from
\begin{equation}
  R_h = \left[ \frac{1}{N^2} \sum_{i \ne j} \frac{1}{r_{ij}} \right]^{-1},
\end{equation}
where $r_{ij}$ is the distance between one pair of beads.
\red{The ratio $R_h/R_g$ is also an indication of the chain shape. 
This ratio attains a value of 1.25 for a self-avoiding chain, and about 2.25 for a stiff rod-like chain \cite{OuZhaoyang2005}. }

\item {\em The asphericity.}
\red{The gyration tensor calculated before contains more information than the average chain size. 
Using the diagonalized gyration tensor, we can compute two shape descriptors that characterize how much the chain shape deviates from a sphere or a cylinder.
One of them is the asphericity,}
\begin{equation}
  b = \lambda_1^2 - \frac{1}{2} (\lambda_2^2 + \lambda_3^2),
\end{equation}
which is a non-negative number. A zero value corresponds to a spherical shape
($\lambda_1^2=\lambda_2^2=\lambda_3^2$), while for a rod-like object one
has $b \approx \lambda_1^2$ ($\lambda_1^2 \gg \lambda_2^2, \lambda_3^2$).

\item {\em The acylindricity.}
\red{The other shape descriptor is the acylindricity,}
\begin{equation}
  c = \lambda_2^2 - \lambda_3^2,
\end{equation}
which is also a non-negative number.  A zero value corresponds to a shape
with uniaxial symmetry, e.g., ellipsoidal or cylindrical
($\lambda_2^2=\lambda_3^2$), i.e., the object appears circular when projected
on to the plane perpendicular to the $\lambda_1$ axis. A large positive value
indicates a deviation from the cylindrical shape.  
\end{enumerate}

\begin{figure}[htbp]
  \centering
  \includegraphics[width=1.0\columnwidth]{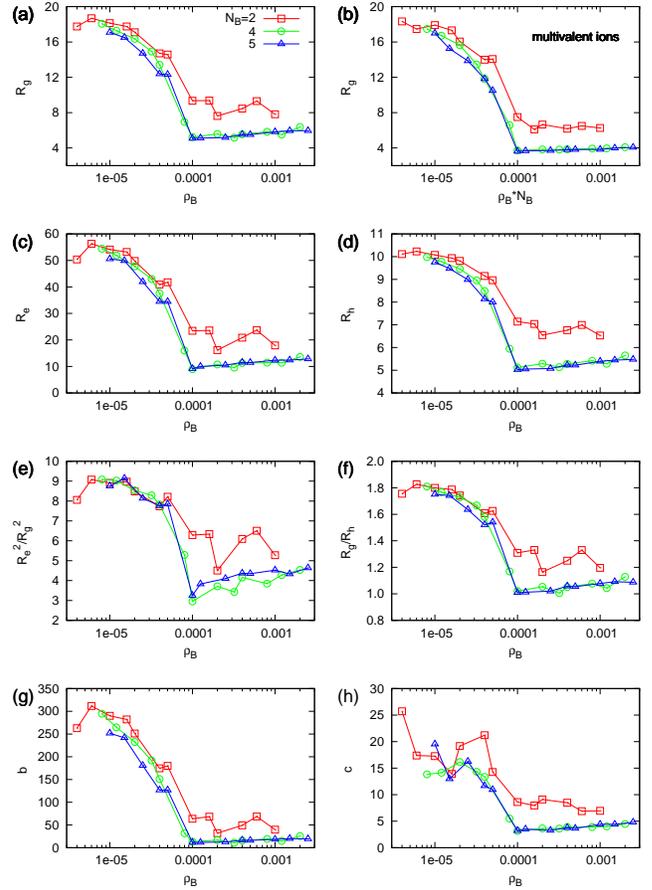}
  \caption{\red{(a) Radius of gyration $R_g$, (c) end-to-end distance $R_e$, (d) hydrodynamic radius $R_h$, (e) the ratio $R_e^2/R_g^2$, (f) the ratio $R_h/R_g$, (g) the asphericity $b$, and (h) the acylindricity $c$, as a function of the B-monomer concentration. 
The radius of gyration for multivalent salts is shown in (b) for comparison.  }
The A-monomer concentration is $10^{-4}\, \sigma^{-3}$.}
  \label{fig:gyr_box1}
\end{figure}

\begin{figure}[htbp]
  \centering
  \includegraphics[width=1.0\columnwidth]{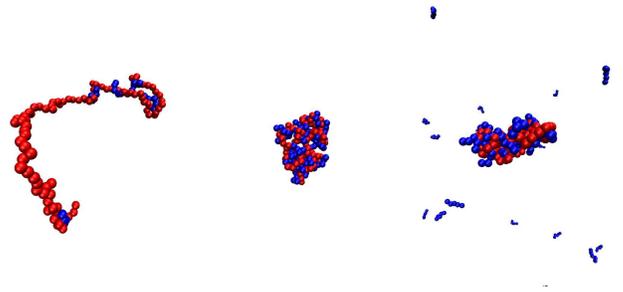}
  \caption{Snapshots of the polyelectrolyte complex at A-monomer
concentration $10^{-4} \, \sigma^{-3}$. For clarity, only the A-chain (red) and
B-chains (blue) are shown. The B-chains have a length of $N_B=5$, and B-monomer
concentrations are $\rho_B = 2.5 \times 10^{-5}, 10^{-4}$, and $2.5 \times
10^{-4}\, \sigma^{-3}$. Visualization is made by VMD \cite{VMD}.}
  \label{fig:snap_box1}
\end{figure}

The time-averaged values of \red{above quantities}
for A-monomer concentration $\rho_A=10^{-4}\, \sigma^{-3}$ are shown in
Fig.~\ref{fig:gyr_box1} as a function of B-monomer concentration.  
Three different length of B-chains are considered, $N_B = 2,4$ and $5$.  In
Fig.~\ref{fig:snap_box1}, representative snapshots of the complex are shown for
$N_B=5$ at different B-monomer concentrations: $\rho_B = 2.5 \times 10^{-5},
10^{-4}$, and $2.5 \times 10^{-4} \, \sigma^{-3}$.

We start with discussing the lowest B-chain concentration, which is \red{close}
to the situation of a free A-chain with only its counterions. Since the
A-chain is strongly charged, the polyelectrolyte assumes an extended form.
The repulsion between like charges on the chain backbone causes the
polyelectrolyte to extend, and the end-to-end distance ($\sim 55\, \sigma$) is
much larger than that of a freely-jointed chain with the same length
($\sqrt{N_A}r_b = 10\, \sigma$).  Nevertheless, the chain conformation is
still far from the rigid-rod limit, which would correspond to an
end-to-end distance $N_A r_b = 100\, \sigma$.  This is partially due to
the counterion condensation.  In our simulation, the Manning parameter
$\Gamma=\ell_B/ \ell_{Q}$, which is the ratio between the Bjerrum length
$\ell_B=3.0\, \sigma$ and the average charge separation on the
polyelectrolyte backbone $\ell_{Q}=r_b=\sigma$, is equal to $3$.  When $\Gamma>1$,
most counterions stay close to the polyelectrolyte backbone, effectively
reducing the repulsion between the charged beads.

When the B-monomer concentration increases, the behavior of the size of the
A-chain reveals two distinct regimes, which are separated by the neutralization
concentration $\rho_B = \rho_A = 10^{-4} \, \sigma^{-3}$.  When positive
B-chains are added to the solution, two main effects take place.  \red{Firstly,
the electrostatic attraction between the A- and B-chains causes an accumulation
of B-chains around the A-chain backbone.  Secondly, the counterions and coions
(monovalent cations and anions) are released from the A-chain backbone and from
the B-chains into the solution, which is associated with a great gain in
entropy.  Both effects favor the complex formation, and the latter appears to
be the main driving force for the polyelectrolyte complexation in highly
charged systems \cite{OuZhaoyang2006}.  Once the complex is formed, the A-chain
size shows a sharp decrease at small B-monomer concentration.  The collapse of
the A-chain is driven by the electrostatic interaction.  \fs{Apart from the
fact} that it is energetically favorable for the complex to have a compact
structure, the B-chains in the complex can also connect two distant monomers on
the A-chain backbone, resulting \fs{in} bridge formation.  This is very similar to
the ion-bridging effect in multivalent salt solution \cite{delaCruz1995}, which
we will discuss in Section \ref{sec:cdn}. }

The complex reaches its minimum size when the concentration of B-monomer and
A-monomer are equal, as all monovalent counterions are replaced by the
B-chains.  Further increasing the B-chain concentration causes the complex
to expand slightly, but the compact globule structure remains. 
\red{The two ratios [Fig.~\ref{fig:gyr_box1}(e) and (f)] and the two shape
descriptors [Fig.~\ref{fig:gyr_box1}(g) and (h)] 
display the same features as the radius of gyration}. 
The A-chain initially has an
elongated shape, and becomes more spherical as the B-monomer
concentration is increased.
For longer B-chains $N_B=4,5$, \red{the characteristic quantities}
of the complex are almost independent of the chain length. This indicates that
the behavior of the complex is indeed dominated by electrostatics and
counterion/coion release, as discussed above, and depends little on the chain
characteristics of the oligocations. In contrast, the shortest oligocations
($N_B=2$) are less effective complexing agents than the longer ones, as evident
from the fact that the size of the complex is larger. For such small values of
$N_B$, the translational entropy of the oligomers becomes important and acts
against complexation.

For comparison, we also show data for the radius of gyration when the
B-chains are replaced by multivalent cations with the same charge
\red{[Fig.~\ref{fig:gyr_box1}(b)]}.  The overall trends are similar, and our results
are consistent with the simulation studies of Hsiao
\cite{HsiaoPai-Yi2006,HsiaoPai-Yi2006a,HsiaoPai-Yi2006b}.  
The size of the complex formed by multivalent cations is slightly smaller than 
that of the complex formed by the oligocations, as the overall volume of the
multivalent cations is smaller.

\begin{figure}[htbp]
  \centering
  \includegraphics[width=1.0\columnwidth]{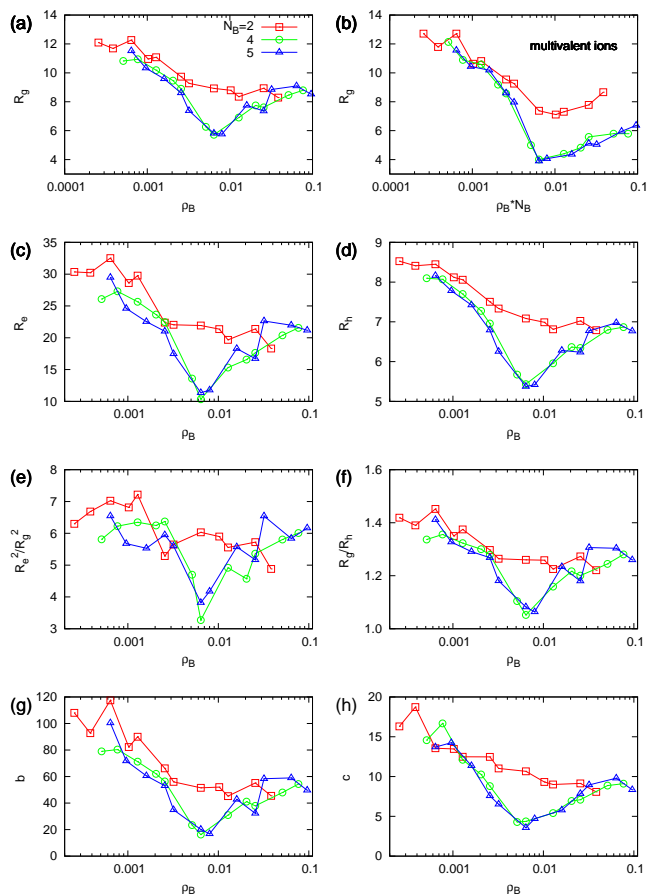}
  \caption{Similar plots as Fig.~\ref{fig:gyr_box1} for A-monomer 
  concentration $\rho_A=6.4 \times 10^{-3}\,\sigma^{-3}$.}
  \label{fig:gyr_box3}
\end{figure}

\begin{figure}[htbp]
  \centering
  \includegraphics[width=1.0\columnwidth]{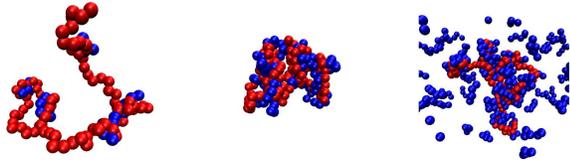}
  \caption{Snapshots of complexes at A-monomer concentration $6.4 \times
10^{-3} \, \sigma^{-3}$. For clarity, only the A-chain (red) and B-chains
(blue) are shown. The B-chains have a length of $N_B=4$, and B-monomer
concentrations are $\rho_B = 1.6 \times 10^{-3}, 6.4 \times 10^{-3}$, and
$1.6\times 10^{-2} \, \sigma^{-3}$. Visualization is made by VMD \cite{VMD}.}
  \label{fig:snap_box3}
\end{figure}

The single-chain conformation also depends on the A-monomer concentration.
Figure \ref{fig:gyr_box3} shows the radii and shape descriptors for a higher
A-monomer concentration $\rho_A=6.4\times 10^{-3}\, \sigma^{-3}$, and
Fig.~\ref{fig:snap_box3} a selection of representative snapshots. As in the
more dilute system (Fig.~\ref{fig:gyr_box1}), one observes a chain collapse
with increasing $\rho_B$ up to the neutralization point where the B-monomer
concentration equals the A-monomer concentration.  However, in contrast to
Fig.~\ref{fig:gyr_box1}, the size of the complex increases significantly when
the B-monomer concentration exceeds $\rho_B=\rho_A$.  All three radii show a
large positive slope as one increases the B-monomer concentration.  The change
of the chain conformation is also apparent when looking at the snapshots \fs{in}
Fig.~\ref{fig:snap_box3}, which show that the A-chain reexpands at high
B-monomer concentration.

This reentrance phenomenon can presumably be explained by the screening
effect of the free ions (counterions, coions, and free oligocations), which are
much more abundant in this system than in the more dilute system of
Fig.~\ref{fig:gyr_box1}. Indeed, if one naively extracts a Debye screening
length from the microion concentrations {\em via} $\lambda_D = \sqrt{1/4 \pi l_B
\sum_c Z_c^2 \rho_c}$ (where the sum $c$ runs over the microion species with
valency $Z_c = 1$), one obtains values between $\lambda_D = 16.3\, \sigma$ (at
$\rho_B \to 0$), $\lambda_D = 11.5\, \sigma$ (at $\rho_B = \rho_A$) up to
$\lambda_D = 3.1\, \sigma$ (at $\rho_B = 2.5 \times 10^{-3}\, \sigma^{-3}$) for
the dilute systems considered in Fig.~\ref{fig:gyr_box1},
hence the Debye screening length always exceeds the Bjerrum length, $\lambda_D
> \ell_B$. In the more concentrated system considered in Fig.~\ref{fig:gyr_box3},
the screening length ranges between $\lambda_D = 2.0\, \sigma$ (at $\rho_B \to 0$),
$\lambda_D = 1.4\, \sigma$ (at $\rho_B = \rho_A$) up to $\lambda_D = 1.1\, \sigma$
(at $\rho_B = 1.6 \times 10^{-2} \, \sigma^{-3}$), which is smaller than the
Bjerrum length. Here, screening clearly becomes important. This is already
apparent from the fact that the characteristic chain radii of bare A-chains (at
$\rho_B \to 0$) are significantly smaller in the concentrated case than in the
dilute case, {\em i.e.}, the electrostatically driven chain stretching is much less
pronounced. Upon adding B-chains, the A-chain collapses both in the dilute and
in the concentrated system. However, both factors driving the complex
formation, the direct electrostatic force between charges as well as the
entropy gain associated with counterion/coion release, are reduced at high
microion concentrations. Hence the oligocations are less tightly bound to the
complex. Beyond the neutralization point, they can more easily change place
with excess oligocations in solution; they can move in and out, the complex
becomes looser and reexpands. It is interesting to note that right at the
neutralization point, the radii and shape factors of the complex are almost the same for the dilute and more concentrated system.

In other respect, the behavior of the A-chain in the more concentrated
system is similar to that of the dilute system. As long as the B-chains are not
too short, the curves for size and shape parameters versus B-monomer
concentration almost lie on top of each other for different chain lengths
$N_B=4,5$.  For short B-chains ($N_B=2$), the minimum of the radii disappears,
and is replaced by a plateau when $\rho_B$ reaches the neutralization
concentration. The reference simulations with multivalent ions show that the
effect of oligocations is very similar to that of multivalent ions with the
same charge.

\subsection{Single-chain structure factor}
\label{sec:soq}

\begin{figure}[htbp]
  \centering
  \includegraphics[width=0.8\columnwidth]{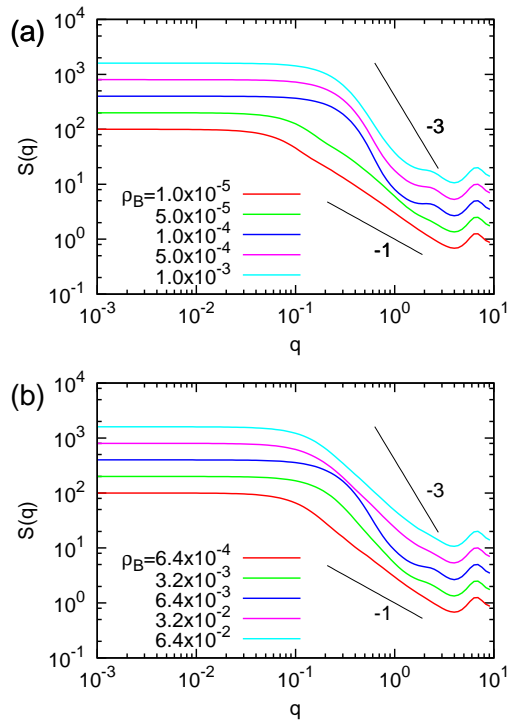}
  \caption{Static structure factor $S(q)$ of the A-chain for A-monomer
concentration (a) $10^{-4}\,\sigma^{-3}$ and (b) $6.4\times10^{-3}\,\sigma^{-3}$.
The B-chains has a length $N_B=5$. For clarity, the $n$th curve in each
graph has been shifted upwards by multiplying $S(q)$ with a factor $2^{n-1}$.}
  \label{fig:soq}
\end{figure}

In this section, we compute the static structure factor of the polyanion chain.
This quantity can be measured in scattering experiments and provides essential
information about the chain scaling factor.  The static structure factor is
defined as 
\begin{equation} 
S(\mathbf{q}) = \frac{1}{N} \sum_{i, j} e^{ i
\mathbf{q} \cdot ( \mathbf{r}_i - \mathbf{r}_j ) } 
\end{equation} 
where $\mathbf{q}$ is the scattering wavevector.  If spherical symmetry can be
assumed, the orientation of the wavevector can be integrated out and the
structure factor only depends on the magnitude of the wavevector.  In
Fig.~\ref{fig:soq}, $S(q)$ is plotted for $N_B=5$ and different $\rho_B$
values.  For clarity, the different curves in the figure have been 
shifted upwards on the y-axis with respect to each other.

The single-chain structure factor reveals useful information about the
characteristic length scales of the polyelectrolyte chain.  In our case, there
are at least two length scales: one is the radius of gyration $R_g$, and the
other one is the monomer size $r_b$.  In the small wavevector limit, $qR_g
\ll 1$, $S(q)$ has a form $N(1-(qR_g)^2/3)$.  In the regime $qr_b > 1$,
the self-scattering of the monomer is the only contribution, and $S(q)$ has a
universal value of \fs{one}.  In the intermediate regime ($R_g^{-1} \ll q \ll
r_b^{-1}$), $S(q)$ shows power-law behavior.  When plotted in a log-log plot,
$S(q)$ has a linear slope in this intermediate regime.  The slope $s$ can be
related to the scaling exponent $\nu$ {\em via} the relation $\nu=-1/s$.  For
rodlike chain, one has $\nu=1$ and $s=-1$, for Gaussian chains, 
$\nu=1/2$ and $s=-2$, and for close-packed globules, $\nu=1/3$ and $s=-3$.

In the dilute systems with A-monomer concentration $10^{-4}\, \sigma^{-3}$
[Fig.~\ref{fig:soq}(a)], the value of the linear slope starts close to
$s=-1$, and changes to a value close to $s=-3$ when the B-monomer
concentration increases.  The change of the scaling factor indicates that
the chain conformation changes from an extended shape to a compact globule
structure.  At high B-monomer concentrations, a small hump appears in $S(q)$ at
around $q=2$.  This indicates that the A-chain is not homogeneously
distributed inside the globule.  The presence of B-chains induces a
certain degree of organization in the A-monomer distribution. In the more
concentrated systems with A-monomer density $6.4\times 10^{-3}\, \sigma^{-3}$
[Fig.~\ref{fig:soq}(b)], the slope takes the value near $s=-1$, changes to
$s=-3$ at intermediate B-monomer concentration, and reverses back to $s=-1$
when the B-chain concentration is high.

\begin{figure}[htbp]
  \centering
  \includegraphics[width=0.8\columnwidth]{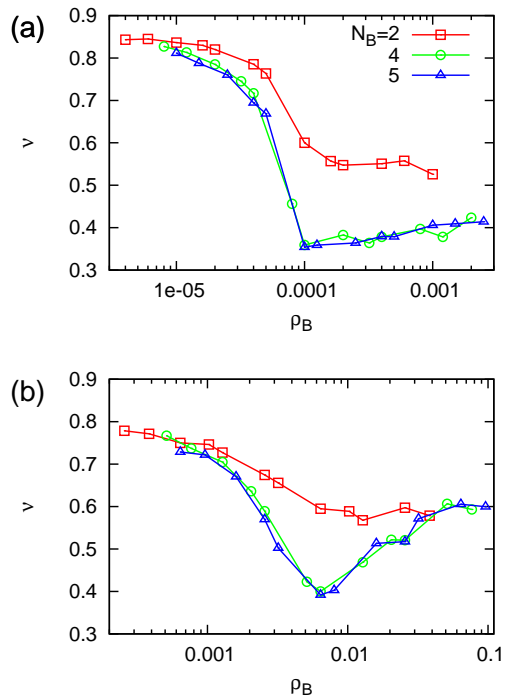}
  \caption{The scaling factor $\nu$ as a function of B-monomer density $\rho_B$
for A-monomer density (a) $10^{-4}\, \sigma^{-3}$ and (b)
$6.4\times10^{-3}\, \sigma^{-3}$. }
  \label{fig:nu}
\end{figure}

We compute the swelling factor $\nu$ by performing linear fits to the log-log
plots of $S(q)$ in the power-law regime.  Figure \ref{fig:nu} shows the
resulting values for $\nu$ as a function of B-monomer concentration for
different B-chain lengths  The behavior of the scaling factor is
similar to that of the radius of gyration, and the physical picture is similar
to what we have described in the last section.  For low A-monomer
concentration, the scaling factor decreases as the B-chain concentration
increases, and reaches a plateau once the neutralization condition is
satisfied.  For high A-monomer concentration, after the neutralization, the
scaling factor starts to increase, indicating a reexpansion of the A-chain.

\subsection{Radial distribution function}
\label{sec:rdf}

\begin{figure}[htbp]
  \centering
  \includegraphics[width=0.8\columnwidth]{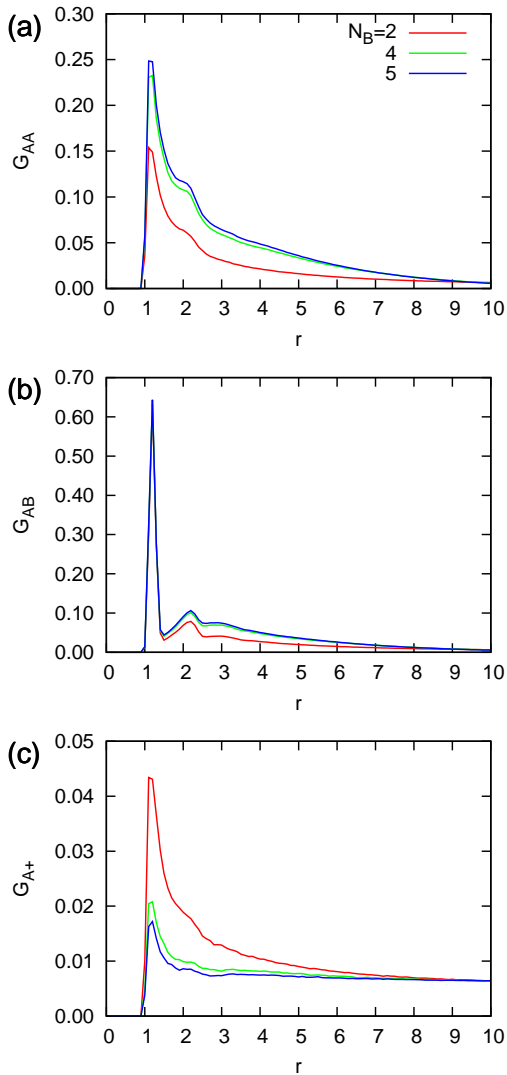}
  \caption{Representative radial distribution function (a) of A-monomers
$G_{AA}(r)$, (b) between A-monomers and B-monomers $G_{AB}(r)$, and
(c) between A-monomers and counterions $G_{A+}(r)$. The plots are
shown for equal A-monomer and B-monomer concentration, 
$\rho_A = \rho_B = 6.4\times 10^{-3}\,\sigma^{-3}$.}
  \label{fig:rdf}
\end{figure}

The complex not only contains the A-chain, but also the oppositely
charged B-chains as well as possibly some counterions and coions. 
Therefore, it is important to understand how the different species are
distributed inside and around the complex.  We first follow the standard
practice to compute the radial distribution functions.  If we choose a type I
bead as the center, the radial distribution function $G_{IJ}(r)$ represents the
number density of type J beads as a function of the distance $r$ to
the central I bead.  We have calculated the radial distribution function for
A-monomers $G_{AA}(r)$, between one pair of A-monomer and B-monomer
$G_{AB}(r)$, and between one pair of A-monomer and counterions $G_{A+}(r)$.
These distribution functions are plotted in Fig.~\ref{fig:rdf} for the
concentration $\rho_A=\rho_B=6.4\times 10^{-3}\, \sigma^{-3}$.  The results for
different B-chain length are also shown.  For different concentrations, the
figures are similar.

The radial distribution functions exhibit relatively little structure, as
is characteristic of a fluid. In Fig.~\ref{fig:rdf}(a), the first peak at
$r=1\,\sigma$ indicates the chain connectivity as the equilibrium bond length is
around unit length.  There is also a weak second peak around $r=2\,\sigma$, but
the correlation smears out at a larger distance, as our polyelectrolyte chain
is flexible.  Figure \ref{fig:rdf}(b) shows that there is a significant
amount of B-monomers in the vicinity of the A-chain.  The first peak
at $r=1\,\sigma$ gives the closest distance between oppositely charged A-
and B-monomers.  A relatively weak second peak indicates a loose
ordering of the second shell. The peaks in the radial distribution
functions get larger when B-chain becomes longer.  However, consistent with the
findings discussed above in Sec.~\ref{sec:Rg}, Fig.~\ref{fig:rdf}(a) and (b)
demonstrates that the dependence on the B-chain length is very weak for longer
chains $N_B=4,5$.  A small amount of counterions accumulates around the
A-monomers for B-chain length $N_B=2$, which can be seen from the small peaks
in Fig.~\ref{fig:rdf}(c).  The peak value reduces when the B-chains become
longer.  We note the scale difference between Fig~\ref{fig:rdf}(b) and
(c), indicating that the B-monomers have a much larger tendency to
accumulate around the A-chain than the monovalent counterions.

\subsection{Effective charge}
\label{sec:eff}

\begin{figure}[tbp]
  \centering
  \includegraphics[width=1.0\columnwidth]{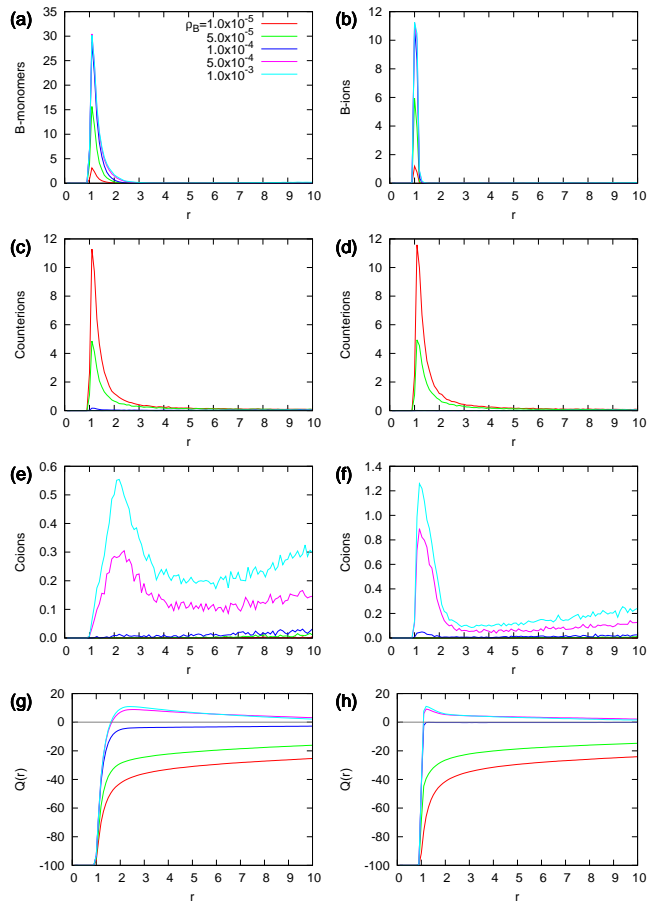}
  \caption{Axial distribution functions of (a) B-monomers, (c) counterions, and
(e) coions. The integrated charge distribution is shown in (g). Here the
A-monomer concentration is $10^{-4}\, \sigma^{-3}$ and the B-chain
length $N_B=5$. The four plots on the left show the oligocation case, and the
right (b,d,f,g) the corresponding results for multivalent salt cations.}
  \label{fig:eff_box1}
\end{figure}

Polyelectrolyte chains have a natural direction along the chain, hence it
is also instructive to analyze the charge distribution with respect
to the polyelectrolyte backbone. Since we consider a flexible polyelectrolyte
in the current study, the backbone of the A-chain is not rigid. 
To determine the charge distribution perpendicular to the backbone, we adopt
a method proposed in Refs.~\cite{Liu2002, Liu2003}: We construct
tubes with different radii $r$ around the A-chain backbone.  For flexible
chains, these tubes do not have a cylinder shape\, but are more
like worm-like tubes. In each snapshot from the simulation, we count
the numbers of various beads between the tubes with radii $r$ and
$r+\Delta r$.  In this study, we choose a bin size of $0.1\,\sigma$. Without
further normalization, we call this quantity the axial distribution
function. One should note that this tube picture is probably not
appropriate when the complex takes a globule structure, but the axial
quantity still provides some information about how charges are distributed in
the direction perpendicular to the backbone.  We show the axial distribution
functions in Fig.~\ref{fig:eff_box1} for A-monomer concentration
$10^{-4}\,\sigma^{-3}$ and B-chain length $N_B=5$.  For comparison, we also
show the same function for the corresponding system containing multivalent
salt cations.

The axial distribution functions show similar features than the radial
distribution functions.  For B-monomers and counterions, they feature
a strong peak at $r=1\,\sigma$.  The peak heights depend on the B-monomer
concentration: for B-monomers [Fig.~\ref{fig:eff_box1}(a)], the peak height
increases as the B-monomer concentration increases, while for counterions
[Fig.~\ref{fig:eff_box1}(c)], it decreases.  This opposite trend
indicates that the highly charged B-chains start to replace the monovalent
counterions in the vicinity of the A-chain backbone as the number of B-chains
in solution increases. Comparing the structure of complexes containing B-chains
with those containing multivalent salt cations, the most pronounced difference
is observed for the axial distribution of the negative coions. In the complexes
containing multivalent salt cations, the anion peak is more pronounced, while
for the case of oligocations, the extension of the coion cloud is broader.

From the axial distribution functions for various charges, we can compute the
integrated charge distribution $Q(r)$, which is defined as the total charge of
particles inside the tube of radius $r$. The results are shown in
Fig.~\ref{fig:eff_box1} (g) and (h). The limiting behavior for small $r$ and
large $r \to \infty$ is similar for all B-monomer concentrations. At very short
distances, $Q(r)$ is equal to $-100$, which is the bare charge of the A-chain.
At large distances, $Q(r)$ approaches zero simply due to the electric
neutrality condition. The behavior at intermediate distances depends on
the concentration of B-chains.  Below the neutralization concentration
($\rho_B<\rho_A = 1.0^{-4}\, \sigma^{-3}$), $Q(r)$ increases monotonically.  The
value of $Q(r)$ rises sharply at first, then slowly increases to its asymptotic
value.  For B-monomer concentrations above the neutralization concentration
($\rho_B> \rho_A = 1.0^{-4}\, \sigma^{-3}$), $Q(r)$ develops a hump at around
$r=2\,\sigma$.  In this case, the positively charged particles bound in the
complex over-compensate the bare negative charge of the A-chain already at
small distances $r > 1.5\, \sigma$. The systems containing multivalent salt
cations show a similar behavior, except that the hump appears close to
$r=1\,\sigma$.

\begin{figure}[tbp]
  \centering
  \includegraphics[width=1.0\columnwidth]{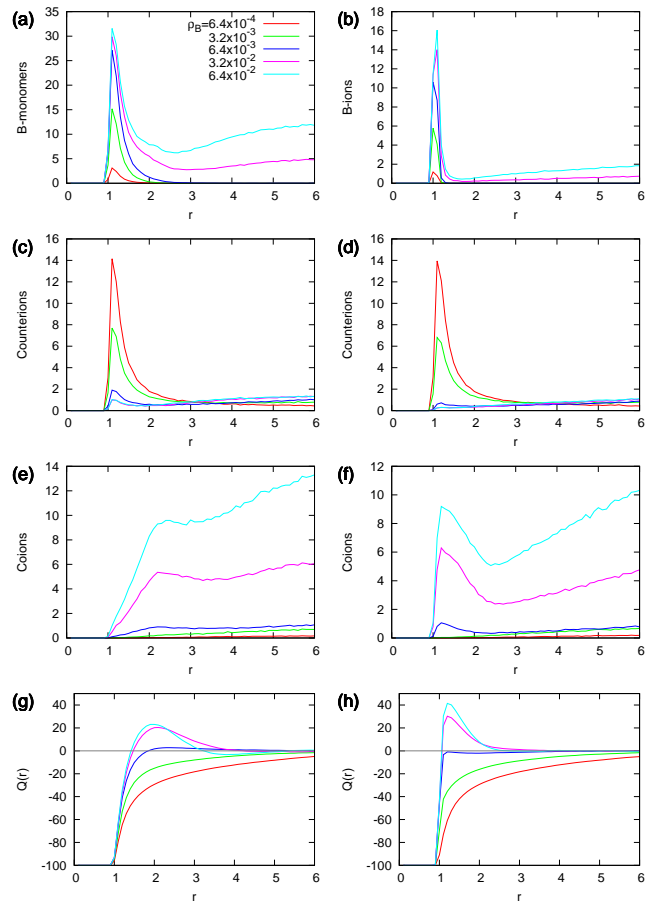}
  \caption{Same as Fig.~\protect\ref{fig:eff_box1} for A-monomer concentration 
$6.4\times 10^{-3}\, \sigma^{-3}$.}
  \label{fig:eff_box3}
\end{figure}

These general features do not change significantly as the A-monomer
concentration is increased.  Fig.~\ref{fig:eff_box3} shows the axial
distribution functions for the systems with A-monomer concentration
concentration $6.4\times 10^{-3}\, \sigma^{-3}$. Compared to the dilute
systems, the main difference is that $Q(r)$ reaches zero much more rapidly, and
that it shows a weakly oscillatory behavior at B-chain concentrations above the
neutralization concentration. An oscillatory behavior of $Q(r)$ was also 
observed by Hsiao in simulations of polyelectrolyte complexation with
concentrated tetravalent salt ions~\cite{HsiaoPai-Yi2006b}. 
Hsiao found that the period of the oscillation matches with the size
of the tetravalent ions, suggesting that they can be related to layering.
This is consistent with our simulations, where the peak of $Q(r)$ is
also broader in the systems containing the (larger) oligocations than in 
the systems containing multivalent salt cations. Hsiao also found
that the structure of $Q(r)$ is more pronounced for tetravalent ions
with larger radii. Here, we observe the opposite effect: Even though the
size of the oligocations is larger than that of multivalent salt cations, 
the peak of $Q(r)$ is smaller. This can be attributed to the flexibility
of the oligocations, and to the fact that their charge is distributed
over several beads.

One may be tempted to determine an effective charge for the complex as the
value of the integrated charge at some cutoff distance $r_c$, $Q_{\rm
eff}=Q(r_c)$. Important physical quantities such as the electrophoretic
mobility of the complex depend on an effective charge. Unfortunately, looking
at the data in Fig.~\ref{fig:eff_box1} (g) and (h), the appropriate choice for
the cutoff is not obvious. This becomes even more critical at higher
concentrations when the $Q(r)$ oscillates around zero. In such cases, 
even determining the sign of the effective charge becomes difficult.

Nevertheless, it seems reasonable to define the charge inside the complex
by choosing a cutoff in the range of the gyration radius $R_g$.  Here, the
situation is relatively clear, and the results do not depend sensitively on the
specific choice of the cutoff.  At B-monomer (or multivalent salt cation)
concentrations below the neutralization condition, the charges on the A-chain
are not yet compensated within the distance $R_g$, hence the complex carries a
negative charge. Above the neutralization point, the charge of the complex
depends on the A-monomer concentration. In the dilute case
[Fig.~\ref{fig:eff_box1} (g) and (h)], the charges on the oligocations (or
multivalent salt cations) enclosed inside the complex exceed the charge of the
A-chain, and the complex is positive.  In the more concentrated case
[Fig.~\ref{fig:eff_box3} (g) and (h)], the charges roughly balance each other
and the complex is neutral.  Hence complexes exposed to a solution with an
excess of oligocations (or multivalent salt cations) tend to fully compensate 
or even overcompensate the charge on the core A-chain, whereas the compensation
is incomplete in a solution which is undersaturated with oligocations 
or multivalent salt cations.

\subsection{Bound ions}
\label{sec:cdn}

\begin{figure}[htbp]
  \centering
  \includegraphics[width=1.0\columnwidth]{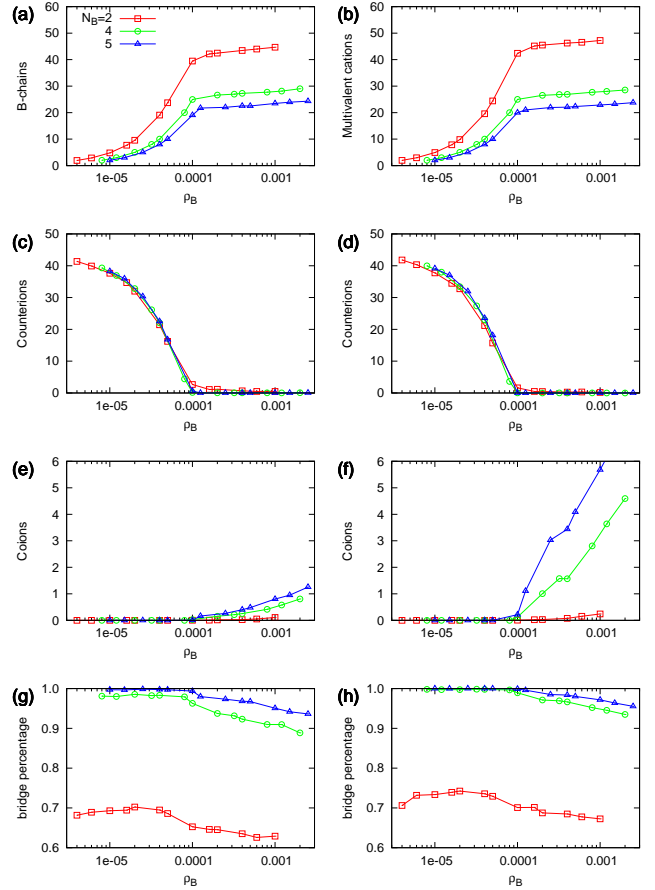}
  \caption{Number of (a) B-chains, (c) counterions, and (e) coions
bound to the A-chain in systems with A-monomer concentration $10^{-4}\, 
\sigma^{-3}$. The graphs in the last row show the percentage of the bridges
(see the main text). The four graphs on the left show the results for
the oligocations case, those on the right the results for multivalent
salt cations.}
  \label{fig:cdn_box1}
\end{figure}

\begin{figure}[htbp]
  \centering
  \includegraphics[width=1.0\columnwidth]{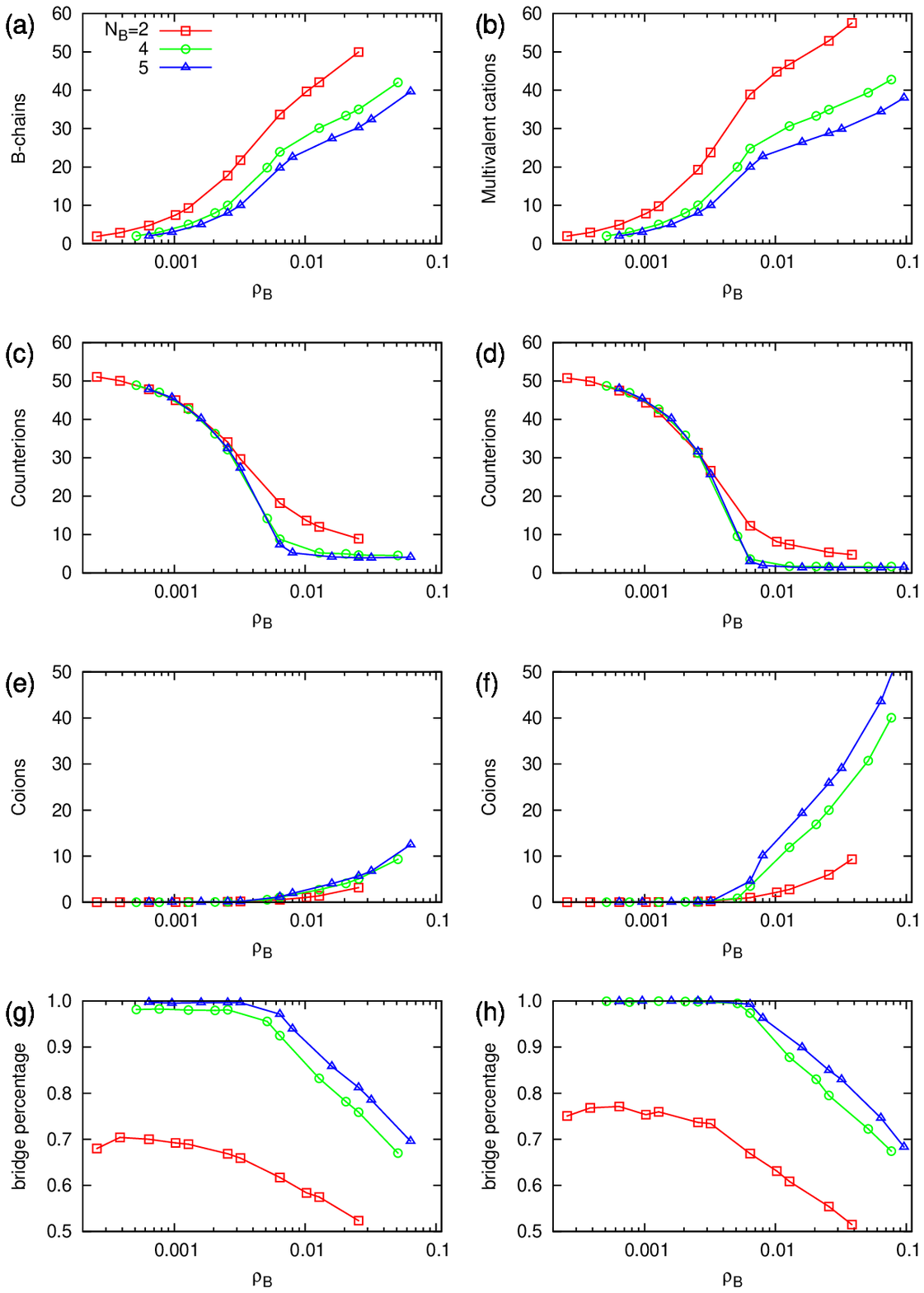}
  \caption{Same as Fig.~\ref{fig:cdn_box1} for A-monomer concentration 
$6.4\times10^{-3}\,\sigma^{-3}$.}
  \label{fig:cdn_box3}
\end{figure}

Whereas assigning an effective charge to the polyelectrolyte complex is a
somewhat delicate issue, as discussed in the previous section, determining the
number of tightly bound ions is much less problematic.
We define a charged bead to be bound to the A-chain if its closest
distance to the A-chain backbone is less than $r_c=1.5\, \sigma$. Similarly, a
B-chain will be called bound if at least one monomer is inside the cutoff
distance.  Figures~\ref{fig:cdn_box1} and \ref{fig:cdn_box3} show the average
number of bound B-chains, counterions, and coions in the complexes, for
$\rho_A=10^{-4}\, \sigma^{-3}$ and $6.4\times 10^{-3}\, \sigma^{-3}$,
respectively.  For comparison, the results for multivalent cations are also
shown in the right panels [Figs.~\ref{fig:cdn_box1} and
\ref{fig:cdn_box3} (b,d,f, and h)].

The number of bound B-chains increases with the B-monomer concentration.
The increase starts gently, but becomes more rapid when the B-monomer
concentration approaches the neutralization point.  Beyond the
neutralization, the number saturates around a fixed value for the dilute
case of $\rho_A=10^{-4}\, \sigma^{-3}$ [Fig.~\ref{fig:cdn_box1}(a)]. The fixed
value is determined by the number of B-chains which is necessary to fully
compensate the negative charge of the A-chain. In the more concentrated
system with $\rho_A=6.4\times 10^{-3}\, \sigma^{-3}$, the number of
bound B-chains continues to increase even beyond the neutralization point.

In contrast, the number of bound counterions decreases with the
B-monomer concentration.  The opposite behavior for counterions and B-chains
can be explained by the release of monovalent counterions upon
complexation, which is favorable due to the gain in translational entropy.
Once the neutralization condition is reached, only very few counterions
remain in close vicinity of the A-chain backbone; most of them have
been replaced by the B-chain. Coions only appear close to A-chain at high
B-monomer concentrations.  Their number increases as the B-monomer
concentration increases.

Finally, we examine the number of ``bridging'' oligocations.  For
multivalent salt cations, de la Cruz \emph{et al.} \cite{delaCruz1995}
have argued that the collapse of the strong polyelectrolyte chain is due
to ion-bridging, \emph{i.e.}, due to the presence of multivalent cation
that connect two distant monomers along the polyanion backbone.  At high
salt concentration, the ion-bridging is screened by the salt, resulting in
a reexpansion of the polyanion chain.  A similar argument has been used to
explain the coil-globule-coil transition of a polymer chain in mixed cosolvents
\cite{Mukherji2014}. In the simulation, we count the number of oligocations
that are within cutoff distance of two non-neighbor A-monomers.  The ratio
between the bridging oligocations and the total number of bound
oligocations is plotted in Fig.~\ref{fig:cdn_box1}(g) and
Fig.~\ref{fig:cdn_box3}(g).  For longer B-chains ($N_B=4,5$), all condensed
B-chains participate in bridge formation below the neutralization
concentration.  When the B-monomer concentration exceeds the neutralization
concentration, the bridge ratio is reduced.  In the dilute systems
with $\rho_A=10^{-4}\, \sigma^{-3}$, the reduction is mild
[Fig.~\ref{fig:cdn_box1}(g)], and this correlates to the small increase of the
A-chain size.  In the more concentrated system with $\rho_A=6.4\times
10^{-3}\, \sigma^{-3}$, the reduction is significant, and the A-chain reexpands
upon increasing B-monomer concentration.  A similar behavior is observed
for the case of multivalent cations.

\section{Summary}
\label{sec:summary}

In this article, we have studied the complex formation of a flexible polyanion
and many short oligocations by Langevin simulations.  We employ a
coarse-grained bead-spring model for the polyelectrolyte chains, and simulate
the small salt ions explicitly.  We consider two different polyanion
concentrations, and systematically vary the oligocation concentration.  We
carefully examine how the oligocations affect the chain conformation,
the static structure factor, the radial and axial distribution of various
charged species, and the amount of ions bound to the polyanion backbone.
We also vary the oligocation length and investigate the influences on the
complex properties.

The long polyanion chain changes its conformation when the shorter oligocations
are added to the solution. Two regimes are observed, which are separated by
the neutralization point where the total charge of the oligocations in the
system exactly matches the charge on the polyanion. Below the neutralization
concentration, the polyanion chain rapidly collapses into to close-packed
globule structure as oligocations are added to the system. In this regime,
the total charge enclosed in the globule is still negative, {\em i.e.}, the
bound counterions and oligocations do not fully compensate the charge on the
polyanion chain. The behavior in the second regime above the
neutralization point depends on the microion concentration. For low
microion concentration (if the Debye screening length is much larger
than Bjerrum length), the globule stays compact, but the sign of the
enclosed charge changes and becomes positive. For higher microion
concentration (Debye length smaller than the Bjerrum length), the globule
reexpands with increasing oligocation concentration and its net charge almost
vanishes.  The reexpansion of the polyanion chain can be correlated with a
reduction of the fraction of oligocations that bridge between different chain
parts.

One particularly noteworthy result of the present study is that the
behavior of the complexes of polyelectrolytes with oppositely charged
oligolytes is very similar to that of polyelectrolytes with oppositely charged
multivalent salt ions. Even more strikingly, it is almost independent of the
length of the oligolytes, as long as they are not too short. It only depends on
the total concentration of opposite charges on oligolytes or multivalent ions
in the solution, compared to the total concentration of polyelectrolyte charge
in the solution. Our simulations thus strongly support the idea that the main
factor driving the complexation and determining the compactness and structure
of the globule is the release of monovalent counterions and coions. The architecture of the multivalent ions seems to be of secondary
importance.

If this is correct, it has interesting consequences. In practical
applications, one often not only needs to know how to bind molecules within a
complex, but also how to release them again. If the dominant force that keeps
the complex together is counterion/coion release, then the relative binding
strength of oligolytes in the complex solely depends on the oligolyte charge:
Oligocations with higher charges will tend to drive out oligocations with
smaller charges from the complex for entropic reasons -- since a smaller number
of highly charged oligocations is necessary to establish the correct charge
distribution.  Peng and Muthukumar \cite{Peng2015} have recently studied how a
longer polyelectrolyte chain can substitute a shorter chain in a complex made
of oppositely charged polyelectrolyte chains of comparable length. In their
case, the substitution is driven by additional counterion release during the
process. Based on our results, we predict that it should be possible to induce
a similar substitution for complexing oligolytes, which is however driven by
the translational entropy of the oligolytes -- the counterion/coion entropy
does not change during the process.

To quote one technologically relevant example, polyelectrolyte complexes
are discussed as potential carriers for siRNA delivery \cite{Leng2009}. siRNA
molecules (small interfering ribonuclease) are very short pieces of double
stranded RNA (about 20-25 nucleotides) which have enormous therapeutic
potential \cite{Whitehead2009, Davis2010}.  In complexes such as discussed in the present paper, they would
take the role of the oligolytes (in this case negatively charged). Once the
carrier has reached its final destination, they could be released most
efficiently -- according to our results -- by exposing the complex to
oligolytes that carry a higher charge than the siRNA. 
The release process is facilitated by the higher concentration of proteins in the cell cytosol. 
Of course, other factors such as the stiffness of the siRNA may play a role that have not been considered in the present study.

The present simulations were carried out in salt-free conditions, i.e., all
microions in the solution were counterions or coions of the polyelectrolyte and
oligolytes in the system. Real systems contain salt ions as well. However, the
main effect of adding salt is to increase the microion concentration and hence
the Debye screening length. We have studied the effect of microion
concentration by changing the size of the simulation box, hence we can also use
our results to discuss the influence of salt on the behavior of the systems. As
discussed above, increasing the microion concentration does not prevent the
complex formation, but it does affect the properties of the complex at
oligolyte concentrations above the neutralization point: The complex becomes
charge neutral, it reexpands upon adding more oligolyte, and the fraction of
bridging oligolytes is reduced.

In the present study, we have considered systems containing one long
polyelectrolyte, and only varied the number of oligocations in the simulation
box.  Therefore, we have imposed the constraint that the complex
contains only one long polyanion chain, thus focusing on the
situation where the polyelectrolyte concentration is very dilute.
In reality, complexes may contain several A-chains, and one expects a
distribution of complexes with different A-chain numbers. To account for this,
it is important not only to understand the structure of one single complex,
but also the interaction between different complexes. Furthermore, we have
only considered flexible polyelectrolyte chain, \fs{whereas} most polyelectrolytes
are semiflexible. Future studies that include the chain rigidity would be
interesting.

\begin{acknowledgments} 
This work was funded by the Deutsche Forschungsgemeinschaft (DFG) 
through the SFB 1066 (project Q1 and A6), and the National Natural 
Science Foundation of China (NSFC) through the Grant No. 21504004.  
Computational resources at  
JGU Mainz (MOGON high performance center) are gratefully acknowledged.
\end{acknowledgments}

\bibliography{complex}

\end{document}